\documentclass[conference,a4paper]{IEEEtran}

\makeatletter
\newcommand{\linebreakand}{%
  \end{@IEEEauthorhalign}
  \hfill\mbox{}\par
  \mbox{}\hfill\begin{@IEEEauthorhalign}
}
\makeatother

\usepackage[T1]{fontenc}
\usepackage[utf8]{inputenc}
\usepackage{lmodern}

\usepackage{graphicx}
\graphicspath{{assets/images/}}
\usepackage{subcaption}

\usepackage{amsmath,amssymb}

\usepackage[colorlinks=true, linkcolor=blue, urlcolor=blue, citecolor=blue, unicode]{hyperref}

\usepackage{xcolor}

\usepackage{listings}

\usepackage{fvextra}
\DefineVerbatimEnvironment{MyVerb}{Verbatim}{breaklines=true}
\usepackage{cprotect}

\usepackage{amsthm}

\usepackage{todonotes}
\usepackage{enumitem}
\usepackage{fancyhdr}
\usepackage{orcidlink}

\usepackage{CJKutf8}
\newcommand{\zh}[1]{\begin{CJK*}{UTF8}{gbsn}#1\end{CJK*}}

\hypersetup{
  pdftitle    = {autowerkstatt4null: An Off-Board-Diagnostics Ecosystem for Car-Workshops},
  pdfauthor   = {Stephan Bökelmann, René Glitza, Meihui Huang, Odin Holmes, Lukas Jakubczyk, Tabea Röthemeyer},
  pdfsubject  = {Overview Paper for autowerkstatt4null},
  pdfkeywords = {AI diagnostics, federated learning, automotive workshops, oscilloscopes, GAIA-X, REST, WebSocket}
}

\begin{document}

\title{autowerkstatt4null: An Off-Board-Diagnostics Ecosystem for Car-Workshops}

\author{

\IEEEauthorblockN{1\textsuperscript{st} Stephan Bökelmann}
\IEEEauthorblockA{\textit{Experimental Hadron Physics} \\
\textit{Ruhr University Bochum}\\
Bochum, Germany \\
sboekelmann@ep1.rub.de \orcidlink{0000-0002-2119-0064}}
\and
\IEEEauthorblockN{2\textsuperscript{nd} René Glitza}
\IEEEauthorblockA{\textit{Institute of Communication Acoustics} \\
\textit{Ruhr University Bochum}\\
Bochum, Germany \\
rene.glitza@rub.de \orcidlink{0009-0002-6437-5912}}
\and
\IEEEauthorblockN{3\textsuperscript{rd} Meihui Huang (\zh{黄美慧})}
\IEEEauthorblockA{\textit{nabla B engineering UG} \\
Bochum, Germany \\
meihui@nabla-b.engineering \orcidlink{0009-0003-3477-9024}}
\linebreakand
\IEEEauthorblockN{4\textsuperscript{th} Odin Holmes}
\IEEEauthorblockA{\textit{Auto-Intern GmbH} \\
Bochum, Germany \\
odin.holmes@auto-intern.de}
\and
\IEEEauthorblockN{5\textsuperscript{th} Lukas Jakubczyk}
\IEEEauthorblockA{\textit{PROLAB Produkt+Produktion} \\
\textit{THGA Bochum}\\
Bochum, Germany \\
lukas.jakubczyk@thga.de \orcidlink{0009-0008-2288-0615}}
\and
\IEEEauthorblockN{6\textsuperscript{th} Tabea Röthemeyer}
\IEEEauthorblockA{\textit{Auto-Intern GmbH} \\
Bochum, Germany \\
tabea.roethemeyer@auto-intern.de \orcidlink{0009-0003-3853-3814}}

}

\maketitle

\pagestyle{fancy}
\renewcommand{\headrulewidth}{0pt}
\renewcommand{\footrulewidth}{0.4pt}
\fancyhf{}

\fancyfoot[L]{\small\textit{autowerkstatt4null: An Off-Board-Diagnostics Ecosystem for Car-Workshops}}
\fancyfoot[R]{\thepage}

\begin{abstract}
This paper presents \textbf{autowerkstatt4null}, a three-year initiative to empower independent automotive workshops with AI-driven, federated diagnostics. 
The project was funded by the \href{https://www.bundeswirtschaftsministerium.de/Navigation/EN/Home/home.html}{German Federal Ministry for Economic Affairs and Climate Action}. 
The project enhanced the initial diagnostic workflow through newly developed technologies and architectural concepts, incorporating real user feedback. 
The resulting demonstrator showcases key outcomes: a modular measurement platform enabling flexible hardware integration, a secure data exchange hub for trusted collaboration, 
asynchronous online diagnostics that decouples analysis from workshop constraints, and a digital academy supporting technician upskilling. 
Together, these innovations demonstrate how independent workshops can access advanced diagnostic capabilities previously limited to manufacturer tools, strengthening competitiveness and sustainability in the sector.
The project repository is available at: \url{https://github.com/nabla-B/paper_aw4null-overview}.

\end{abstract}


\section{Introduction}
\subsection{State of Car Diagnostics}
\begin{figure}[ht]
  \centering
  \includegraphics[width=0.9\linewidth]{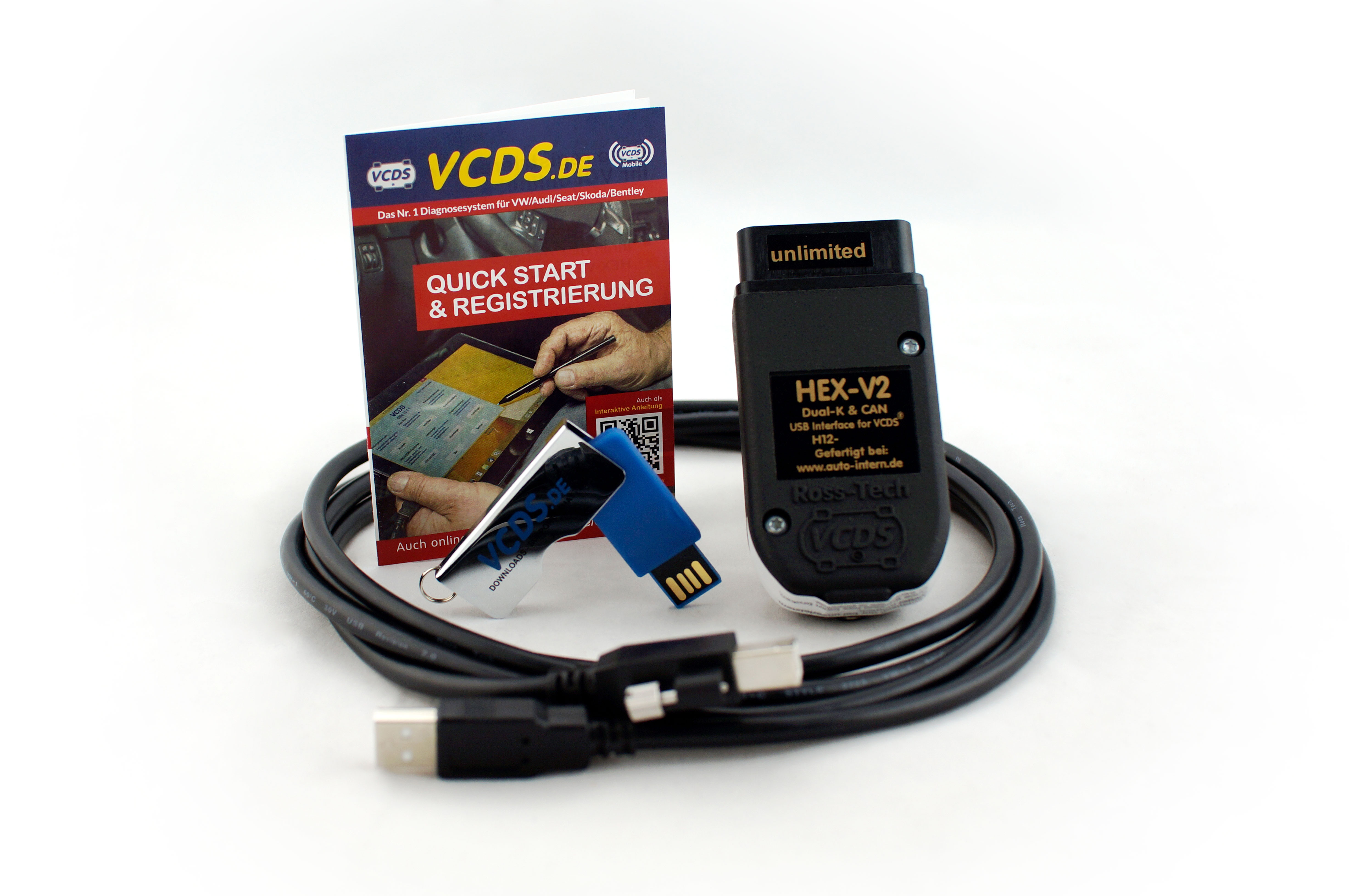}
  \caption{Picture of an OBD device}
  \label{fig:obd}
\end{figure}
Modern vehicles integrate numerous interconnected subsystems, including mechanical components like engines and brakes, 
electronic control units managing sensors and actuators, and electromechanical parts such as electronically controlled valves and motors. 
Market pressure, regulations, and user demands regarding energy efficiency, comfort and passenger safety lead to an ever-increasing complexity in the systems themselves and their interactions.
Diagnostic tools must therefore analyze data across these diverse systems, such as engine management, braking systems, airbag controllers, 
and advanced driver-assistance features, to effectively identify the root cause of the malfunction.

Design goals for these car diagnostic systems are multiple:
\begin{itemize}
  \item Easy to interpret error information delivery
  \item Hardware design with minimal physical controls
  \item Accessible software interface with minimal user interaction
  \item Easy to train and multilingual support
  \item Robustness against physical damage
  \item Cost-effectiveness setup and operation
  \item Minimization of misdiagnosis and repair errors
\end{itemize}
Since some faults can lead to degraded combustion behaviors without immediately noticeable effects for the car’s operator, 
many diagnostic systems—particularly those related to emissions—are legally required to be permanently integrated into the vehicle~\cite{obdregulations, epaobd}. 
These legal requirements are enforced through regulations such as the European Union’s Euro 6 standards and the U.S. Environmental Protection Agency on-board diagnostics (OBD) requirements. 
As a result, a distinction arises between OBD, which must be present in every vehicle by law, and off-board diagnostics, 
which are operated externally in workshops to support more detailed or non-regulated diagnostics.
According to this very definition, OBD tools get delivered with each and every individual car itself, hence adding a) more costs; b) more weight; and c) more complexity to the car.

It is therefore that auto manufacturers are trying to reduce the amount and extent of these systems. 
This creates a tension between manufacturers, who aim to minimize on-board systems, and regulators, who enforce minimum diagnostic capabilities to ensure vehicle safety and environmental compliance.
In opposition to that, off-board systems do not need to be built into every car, but need to be acquired by workshops. 
The market pressure for those tools arises out of the situation that mechanics get paid for maintaining and repairing a car but not necessarily the diagnostics itself. 
That’s why the evaluation of the cost-benefit ratios looks different for these tools.

When a diagnostic need arises in an auto workshop, technicians typically rely on specialized OBD-readout interfaces - in order to read out the fault-codes. 
These tools connect to the vehicle’s onboard systems via standardized interfaces such as the OBD-II port (see Figure~\ref{fig:obd} for an example device). 
Commonly used diagnostic devices include scan tools and code readers from vendors like Bosch, Snap-on, Autel, and Launch. 
These tools allow workshops to read error codes, perform system tests, and monitor sensor data to pinpoint faults. 
Although they are regarded as effective \footnote{mechanics report faster and more precise electrical diagnostics compared to basic 
multimeters\cite{redditscopeefficiency}} they can be expensive and often require frequent updates to support proprietary and evolving communication protocols, 
such as multiple OBD-II variants (SAE J1850, ISO 9141-2, ISO 15765/CAN) and manufacturer-specific protocols like UDS, which are commonly accessed by third-party 
diagnostic tools such as VCDS\cite{obd2protocols, obd2wiki, vcds}.
An essential off-board diagnostic tool for electrical fault analysis is the oscilloscope, 
which has been used in its traditional and auto shop specific variations by car technicians since the late 1950s in Germany~\cite{hameghistory} and became widespread internationally during the 1960s~\cite{autoscopehist}.

\subsection{Future Challenges in ICEV and BEV}
Diversification in the set of vehicles—driven by different models, manufacturers, and subsystem configurations—has led to increased diagnostic complexity, 
requiring varied tools, protocols, and specialist knowledge~\cite{autodiagnosticsdiversity, triddiagnosticcomplexity}. 
One particular fork in the road appears to be the further differentiation between internal combustion engine vehicles – ICEV; and battery electric vehicles – BEV. 
Both technologies confront auto shops with different diagnostic challenges\footnote{See the curated list of articles and sources on this topic at \href{https://gist.github.com/kathamatician/fbc405fd53297b142fcf41163fad2d1e}{this GitHub Gist}},  
yet both have in common that they include electrical signals and purely mechanical parts. 
While fault diagnostics of the purely mechanical parts of the car, e.g. suspension, hinges, brakes, and fenders, can be categorized as a traditional craft, 
the diagnostics of peculiar electrical signals is as diverse as the ever-increasing amount of electrical assistant systems. 

\subsection{Interesting Technologies for Upcoming Car Diagnostic Systems}
Oscilloscopes are among the few tools that allow automotive technicians to observe real-time electrical behavior-something static diagnostic codes alone cannot reveal. 
They are used to pinpoint faults like voltage dropouts, transient glitches, or signal distortion, especially in sensors and control lines.
By capturing these short-lived anomalies, high-speed tools enable more precise fault tracing and reduce trial-and-error repairs. 
This makes them indispensable for diagnosing elusive issues that don’t result in stored diagnostic trouble codes (DTCs)~\cite{tektronix2024, fleetmaintenance2024, picoauto2023, fleetmaintenance2020}.

\begin{figure}[ht]
  \centering

  \begin{subfigure}[b]{0.9\linewidth}
    \centering
    \includegraphics[width=\linewidth]{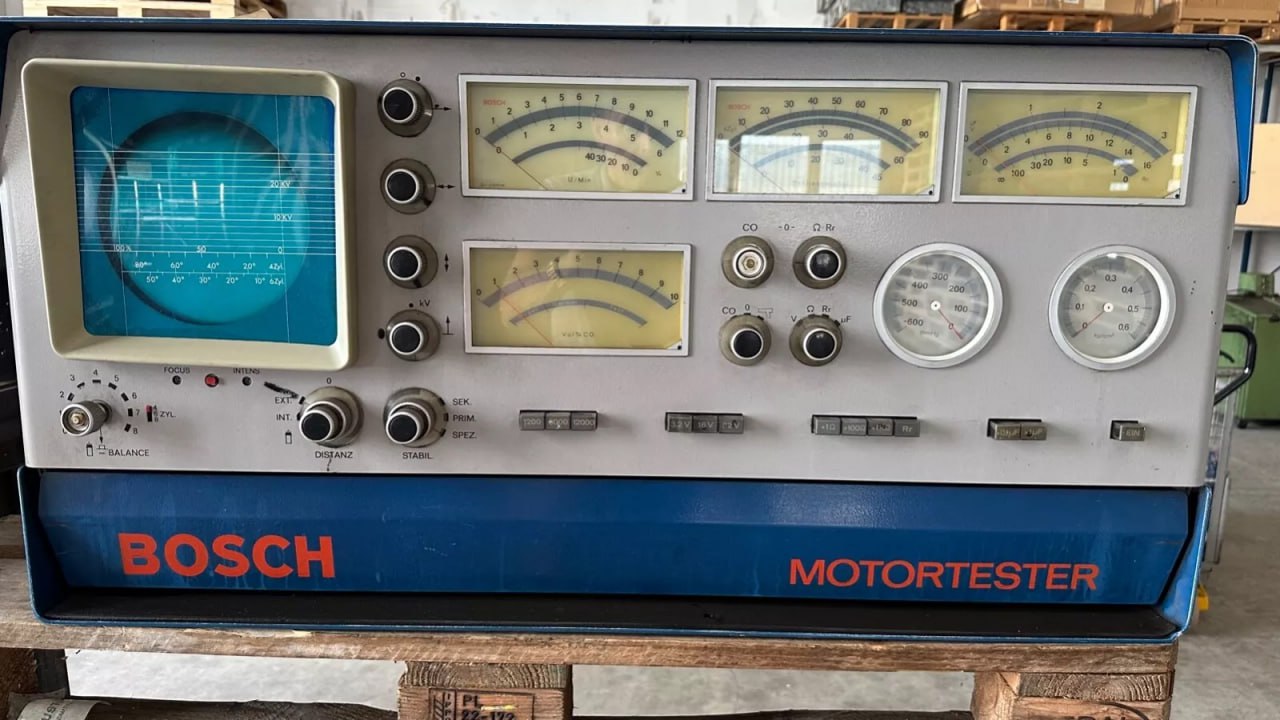}
    \caption{\mbox{Legacy Bosch Motorentester including Oscilloscope}}
  \end{subfigure}

  \vspace{0.5cm}

  \begin{subfigure}[b]{0.9\linewidth}
    \centering
    \includegraphics[width=\linewidth]{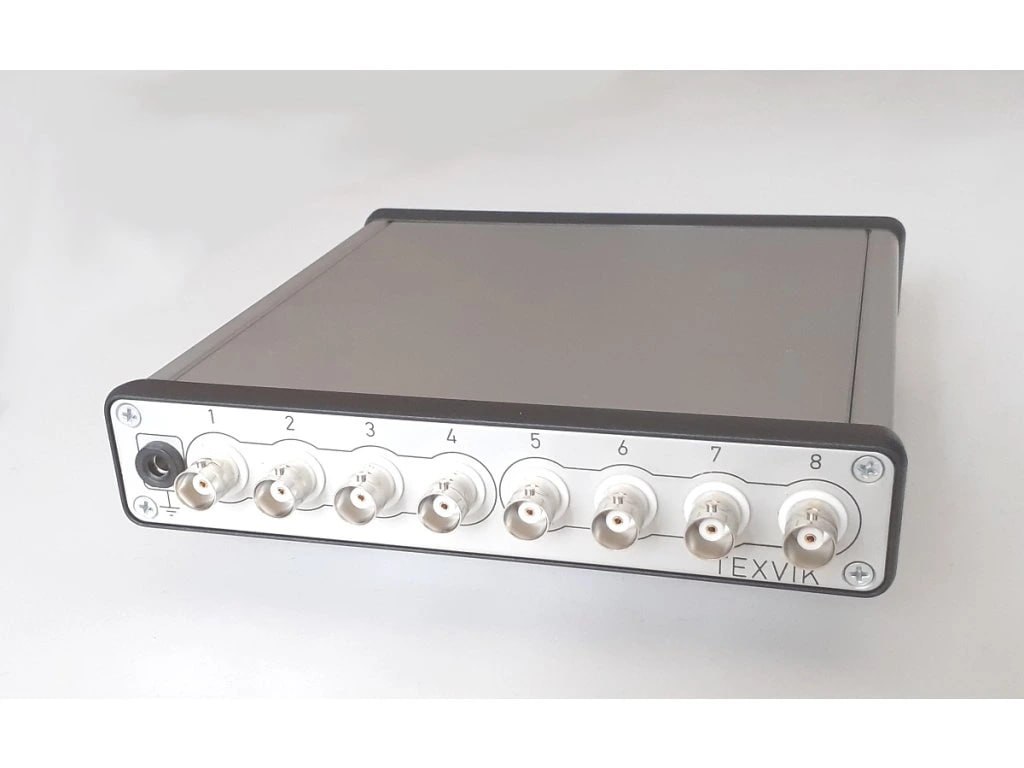}
    \caption{\mbox{Texvik DS8 Mk II v.2 Automotive Oscilloscope}}
  \end{subfigure}

  \vspace{0.5cm}

  \begin{subfigure}[b]{0.9\linewidth}
    \centering
    \includegraphics[width=\linewidth]{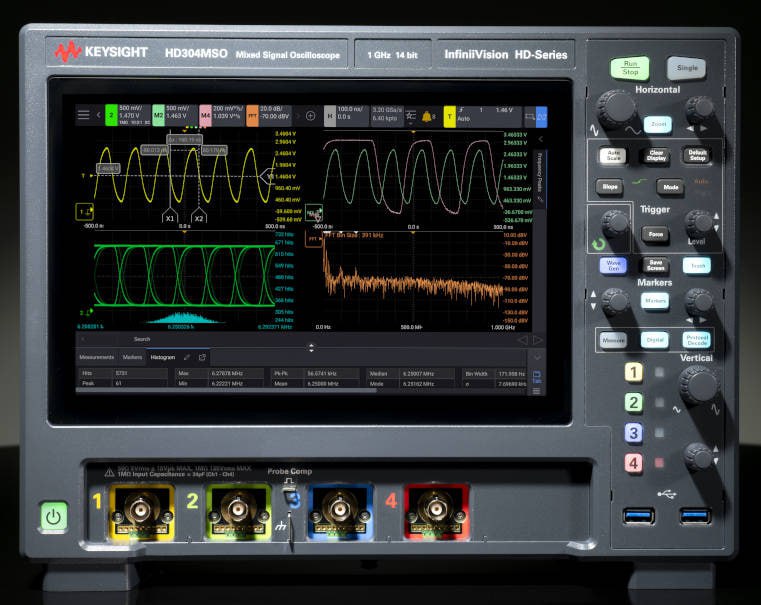}
    \caption{\mbox{Professional Keysight Laboratory Oscilloscope}}
  \end{subfigure}

  \caption{Three different Oscilloscopes}
  \label{fig:scopes}
\end{figure}

Different models of oscilloscopes vary in complexity and features, as shown in Figure~\ref{fig:scopes}, 
which presents examples ranging from a traditional analog scope to more advanced digital and auto-scopes.

By analyzing the dynamic voltage behavior shown on the oscilloscope, technicians can identify engine faults that remain undetectable using current on-board and off-board diagnostic systems.
Moreover, oscilloscopes offer a brand-independent, potentially cost-efficient diagnostic tool—strengthening the earlier argument about the cost-benefit ratio of external diagnostic platforms.

However, effective oscilloscope usage requires electrical knowledge, correct probe handling, and experience in waveform interpretation~\cite{rohde2024, autoditex2023, pico2021, mechanic2024}. 
These barriers make oscilloscopes impractical for many general-purpose workshops. 
As a result, only a subset of specialists can currently harness their full potential. 
Reiterating the previous argument about cost-benefit ratio of off-board diagnostic systems, 
this leads to a situation in which a useful tool cannot be trivially deployed to non-prepared auto shops, due to extensive learning curves.

Now, with the availability of modern microcontrollers, graphical user interface (GUI)-driven workflows, and automated analysis powered by machine learning, 
we aim to reduce the learning curve and increase the accessibility of oscilloscope diagnostics. 
We suspect to enhance teachability and robustness of these tools by employing certain technologies:
\begin{enumerate}[label=\alph*)]
  \item Machine Learning for Signal Clustering
  \item Modern Microcontroller Technology
  \item REST API \& Web Sockets
\end{enumerate}

\vspace{1em}
\subsubsection{Machine Learning for Signal Clustering}
\begin{figure}[ht]
  \centering
  
  \begin{subfigure}[b]{0.9\linewidth}
    \includegraphics[width=\linewidth]{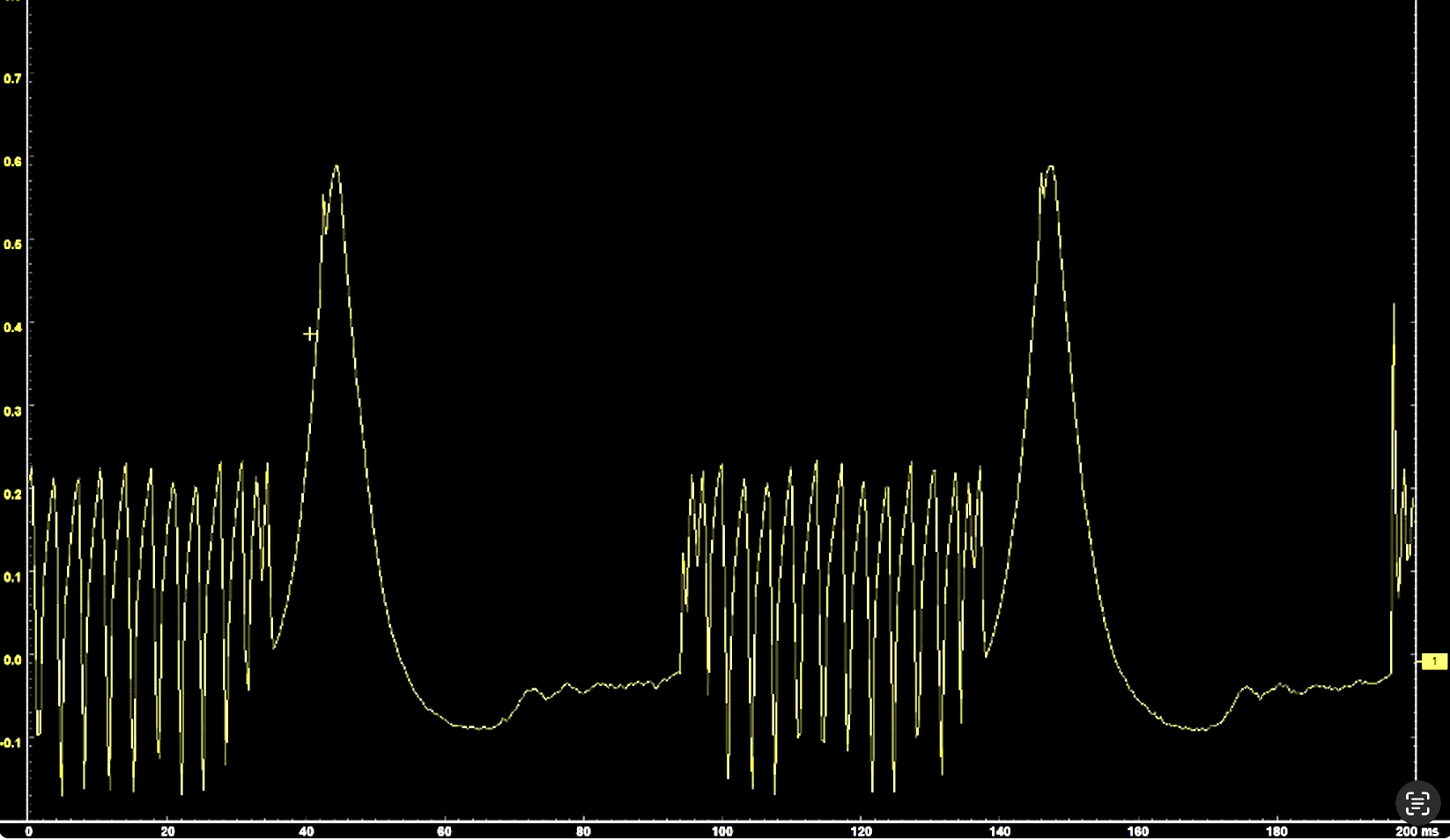}
    \caption{}
  \end{subfigure}
  \vspace{0.5em}
  \begin{subfigure}[b]{0.9\linewidth}
    \includegraphics[width=\linewidth]{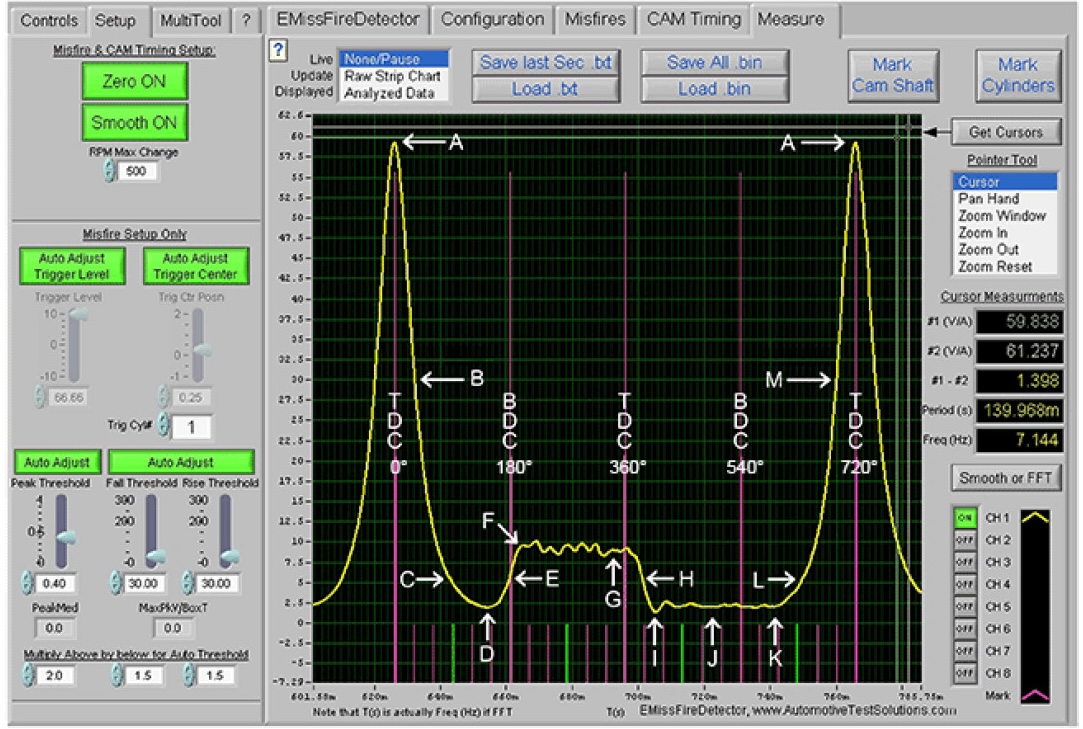}
    \caption{}
  \end{subfigure}

  \caption{Two waveforms: a poor (a) and a good (b) a compression cycle. This can be assumed from the voltage drop while compressing and the following rebound}
  \label{fig:waveforms_good_bad}
\end{figure}
Programed computer functions can be understood as mappings of a domain to a codomain, where the domain is vulgarly referred to as input data and the codomain as output data~\cite{knuth71, wirth76}. 
During the 1950s, researchers at Bell Laboratories and the Institute for Advanced Study in Princeton proposed that such programmatic functions could be executed autonomously by computer systems themselves 
- an idea that anticipated what is now formalized as the Universal Approximation Theorem  \cite{Cybenko1989, Hornik1991}. 
Since then, this niche idea has evolved into a scientific discipline called Artificial Intelligence and Machine Learning \cite{RussellNorvig2020, Goodfellow2016}. 
Latest advances in this area did demonstrate useful opportunities \cite{LeCun2015, Silver2016}. 
It can thus be imagined to define an abstract function that maps a multidimensional time dependent vector of electrical quantities, also known as time series, to a certain fault category. 
Figure~\ref{fig:waveforms_good_bad} shows two example waveforms—one from a poorly and one from a properly functioning compression cycle—highlighting the type of signal variation that machine learning models can be trained to classify.
From the toolkit of machine learning, multiple technologies could be interesting from a research stand point to generate such a function or group of functions, though all having in common a need for training data. 
Therefore we claim, that by automating signal interpretation, the implementation of such functions can reduce the need for manual waveform analysis, 
lowering the level of expertise required and thus decreasing the amount of technician training necessary-while improving diagnostic reproducibility.

\subsubsection{Modern Microcontroller Technology}
Traditional oscilloscopes usually come as all-in-one boxes, with high-accuracy analog front-ends and delicate signal processing units to display the measured voltages on a built-in screen. 
All of the mentioned parts are usually not meant for rough environments such as auto workshops. 
Specially designed electronic oscilloscopes that come in ruggedized cases may end up costing multiple thousands of euros, such as:
\begin{itemize}
    \item PicoScope
    \item FCD Scope
    \item Bosch
    \item Hella Gutmann
    \item Tektronix 1100 Series
    \item Fluke 120B / 190 Series
    \item Keysight U125x / U160x
\end{itemize}

So far, regular microcontrollers have not been used widely in oscilloscopes. 
However, recent generations of microcontrollers offer sampling speeds in the low to mid MHz range, 
sufficient for capturing many automotive signals—such as ignition pulses (typically up to a few hundred kilohertz) or communication protocols like CAN bus (500 kHz). 
For instance, the RP2040’s 12-bit ADC runs up to 500 kS/s~\cite{rp2040ds}, STM32F7 ADC clocks can reach 50 MHz enabling multi‑MS/s operation~\cite{stm32f7ds}, 
and automotive‑grade Renesas RH850/C1M‑A cores operate at up to 320 MHz for high‑speed signal processing~\cite{rh850ds}. 
Typical fault-related signals in cars tend to lie well below 1 MHz, with transient events often occurring on the order of microseconds to milliseconds.

\begin{figure*}[h]
  \centering
  \includegraphics[width=0.9\textwidth]{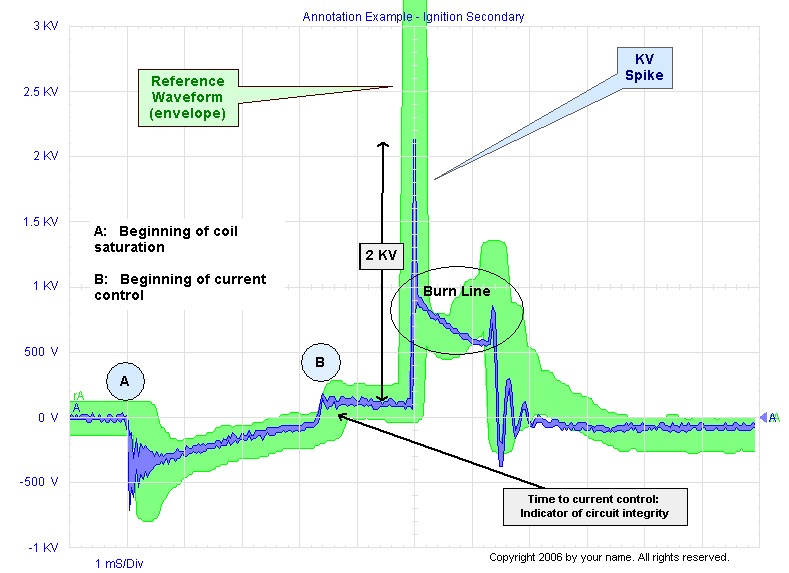}
  \caption{Oscilloscope capture of an ignition secondary waveform showing coil saturation, current control, burn line, and high-voltage spike. 
  Image adapted from a user-contributed post on Electronics StackExchange~\cite{stackexchangeignition}.}
  \label{fig:ignition-scope}
\end{figure*}

Given these characteristics, we claim that microcontrollers capable of sampling at 1–5 MHz with adequate memory for capturing durations ranging from 
milliseconds to seconds could provide meaningful diagnostic data for many use cases in auto shops. 
As illustrated in Figure~\ref{fig:ignition-scope}, typical ignition pulses last for several milliseconds and contain features observable at sampling rates of around 1 MHz.
Compared to faster but more expensive alternatives like FPGAs, developing a sampling oscilloscope on a microcontroller basis with a USB interface to a regular computer, 
tablet, or smartphone could prove a viable way forward toward more robust yet inexpensive diagnostic tools tailored for the workshop environment.

\subsubsection{REST API and Web Sockets}
\begin{figure}[ht]
  \centering
  \includegraphics[width=0.9\linewidth]{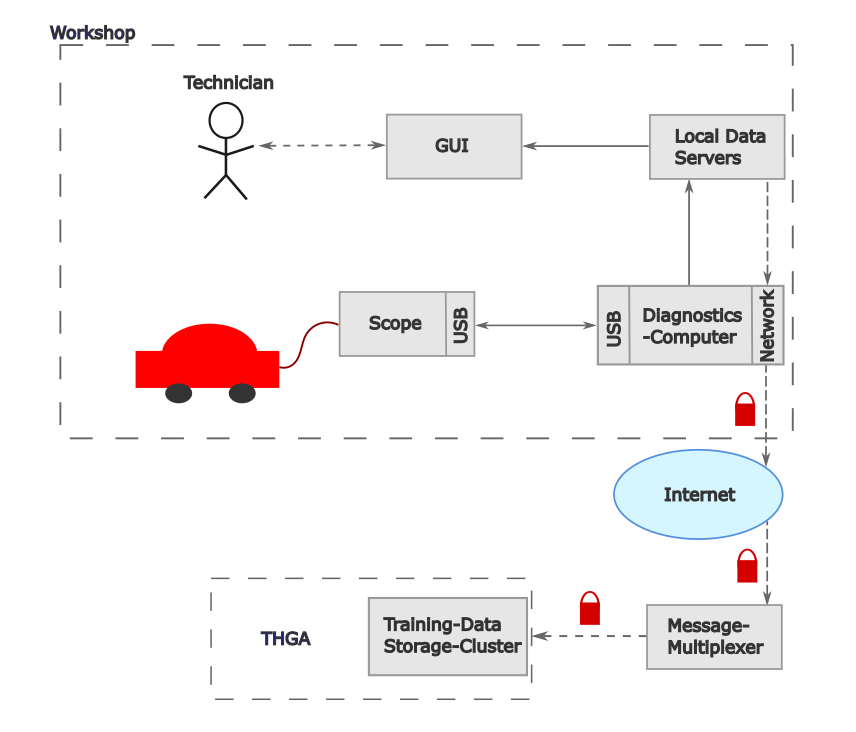}
  \caption{Data-server architecture for oscilloscope integration.}
  \label{fig:data-server}
\end{figure}
Since the two cornerstones - machine learning and microcontroller - based oscilloscopes - have been identified by now, the challenge of connecting them has to be addressed.

Since Machine Learning models require certain amounts of training data as well as centralized compute power~\cite{Verbraeken2020DMLS}, 
the according analysis models can be more easily run on centralized servers than on individual auto shop workstations. 
Yet, the data acquisition with the microcontroller oscilloscopes has to be performed on-site at the auto shop. 
Connecting these two modules could be done in a multitude of ways. 
A conceptual overview of this architecture is illustrated in Figure~\ref{fig:data-server}, where an oscilloscope is connected to a local diagnostics-computer via USB.
This captures the data, memorizes it until a measurement is finished; packs it into a HTTP-message and sends it to a remote machine, which provides access to analysis services and further storage.
The remote machine is answering with a diagnostic message of the captured signal.

With the increasing employment of web technologies -especially Hypertext Transfer Protocol (HTTP)- it can be assumed that handling the data from acquisition to analysis could be implemented in an open-source approach. 
While doing that, the fundamental architecture needs to encompass two different modes of communication: asynchronous and synchronous.

\paragraph{Asynchronous Communication}
Asynchronous communication is not dependent on both ends of a communication channel having the same concept of time. 
This results in a certain pattern of communication that can also be described as a request–response model. 
One common approach to implement this model on the web is through REST (Representational State Transfer), an architectural style for designing networked applications. 
HTTP (Hypertext Transfer Protocol) is a protocol that implements RESTful communication by defining how requests and responses are formatted and exchanged \cite{Fielding2000}.

These communication methods can be used to send a structured package of acquired data, including a recorded time series of a voltage as well as relevant metadata about this measurement, 
such as mileage, sensor type, make-and-model of a car to a server interface called API. 
The server then has a certain amount of time to analyze the posted data, compare it against the internal database, formulate a response with a diagnosis and send it back to the requesting client.

\paragraph{Synchronous Communication}
The described pattern of analysis requires all of the posted data to be already in the RAM of the client machine, while the data itself can only be acquired one sample at a time over a certain period. 
Collecting these individual samples into a structured block necessitates that input channel of the computer as well as the data collecting module run synchronously, sharing the same time reference. 
This can be done via the WebSocket protocol, an extension of HTTP.

We conclude that the middleware between USB-connected oscilloscopes and internet connected analysis service can be implemented using off-the-shelf web frameworks.

\section{Proposed System}
\subsection{Overview}
\begin{figure*}[t]
  \centering
  \includegraphics[width=\textwidth]{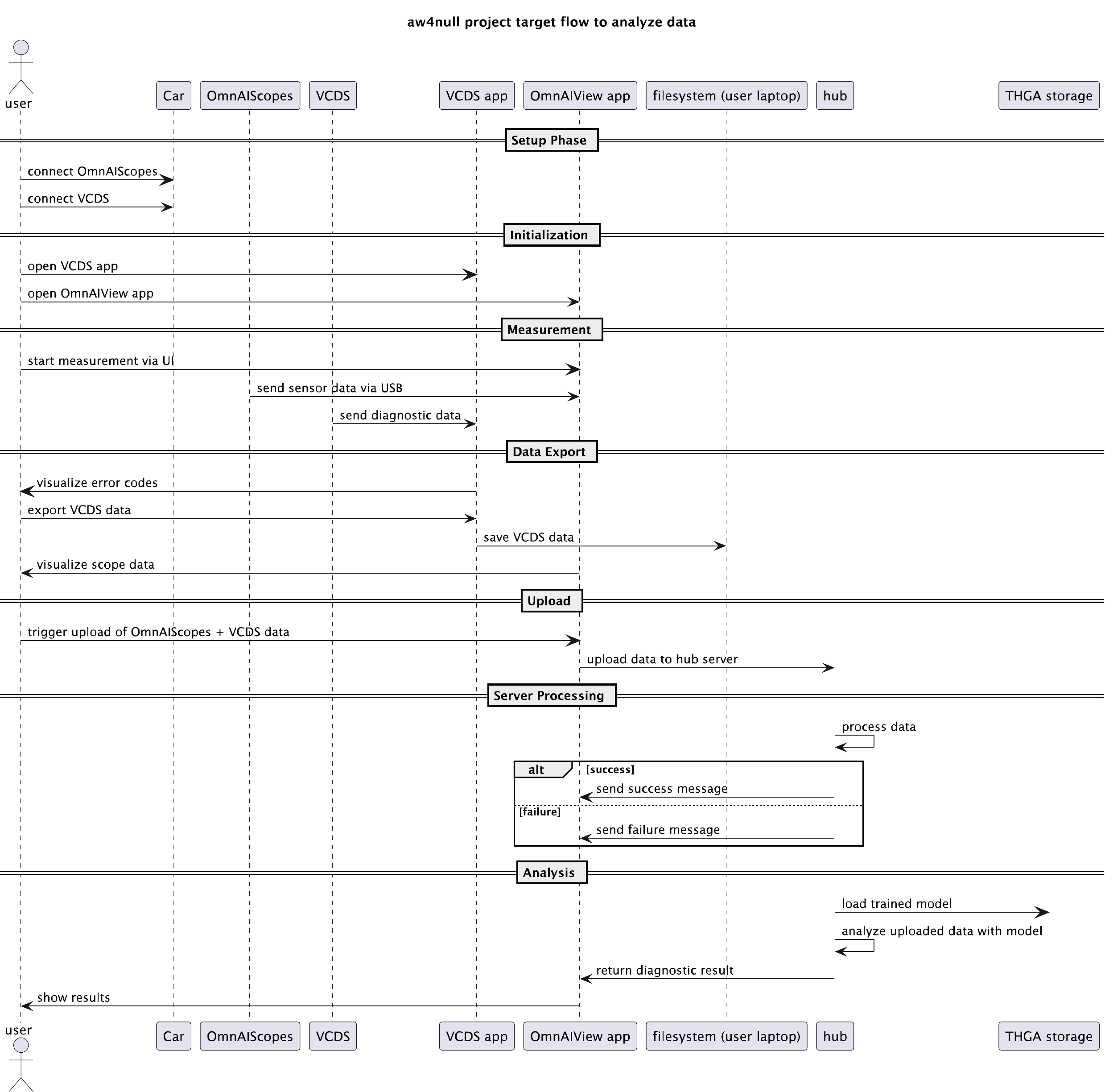}
  \caption{Federated ecosystem architecture: data flow from scopes to AI services to technician UI}
  \label{fig:system_architecture}
\end{figure*}
Based on these previous considerations and the evergrowing complexity within cars, on-board-diagnostic solutions don't necessarily capture all the required use cases.
In this paragraph, a proposed off-board-diagnostic-solution is outlined, which might improve single car shops diagnostics capabilities by employing machine learning and data analytics. 
This system is illustrated in Figure~\ref{fig:system_architecture}, which shows how data flows from oscilloscopes to centralized AI services and back to technicians via a user interface.
Since the central entity of off-board diagnostics is ususally a techinician within a workshop, the first line of argumentation, that is entertained is their perspective on such a system.

\paragraph{Technician Workflow}
A customer's car enters a workshop. 
The first step taken in some cases is the full on-board-diagnostic scan resulting in a protocol of the quiry of all in the car available engine control units (ECU). 
This protocol encapsulates the status in which the car enters the workshop, compiling information such as milage, ECU series numbers, installed system with according hardware and software versions as well as potentially stored DTCs.
It is this protocol that can hint the technician to necessary maintenance or repair actions.
Unfortunately, due to the lack of sophistication of some of the on-board-diagnostic systems, DTCs don't always point out a problem in the most obvious way. 
An example for this is the diagnostic trouble code shown in Figure~\ref{fig:dtc_codes}.

\lstset{
    basicstyle=\ttfamily\small,
    breaklines=true,
    frame=single,
    backgroundcolor=\color{gray!10},
}

\begin{figure*}[ht]
\centering
\begin{lstlisting}[literate={°}{{\textdegree}}1]
    7380 - Manifold Pressure / Boost Sensor (G31)
    P0236 00 [165] - Implausible Signal
    MIL ON - Not Confirmed - Tested Since Memory Clear
       Freeze Frame:
              Fault Status: 00000001
              Fault Priority: 2
              Fault Frequency: 1
              Mileage: 313662 km
              Date: 2014.14.30
              Time: 00:11:44
    
              Engine RPM: 0.00 /min
              Normed load value: 0.0 %
              Vehicle speed: 0 km/h
              Coolant temperature: 30 °C
              Intake air temperature: 19 °C
              Ambient air pressure: 980 mbar
              Voltage terminal 30: 12.762 V
              Unlearning counter according OBD: 40
              Charge air pressure: actual value-MAP_MMV: 1154.71 hPa
              Charge air pressure: specified value-MAP_SP_MMV: 1154.71 hPa
              Ambient air pressure: 98 kPa abs
              Particle filter: difference pressure-DIP_PF: 0.00 hPa
    
    7382 - Manifold Pressure / Boost Sensor (G31)
    P0236 00 [096] - Implausible Signal
    Intermittent - Not Confirmed - Tested Since Memory Clear
       Freeze Frame:
              Fault Status: 00000001
              Fault Priority: 2
              Fault Frequency: 1
              Date: 2000.00.00
              Time: 00:00:00
    
              Engine RPM: 0.00 /min
              Normed load value: 0.0 %
              Vehicle speed: 255 km/h
              Coolant temperature: 34 °C
              Intake air temperature: 19 °C
              Ambient air pressure: 980 mbar
              Voltage terminal 30: 12.536 V
              Unlearning counter according OBD: 40
              Charge air pressure: actual value-MAP_MMV: 1153.47 hPa
              Charge air pressure: specified value-MAP_SP_MMV: 1153.47 hPa
              Ambient air pressure: 98 kPa abs
              Particle filter: difference pressure-DIP_PF: 0.00 hPa
\end{lstlisting}
\caption{Diagnostic trouble codes (DTCs) for the manifold pressure sensor.}
\label{fig:dtc_codes}
\end{figure*}

\textit{Pressure sensor output out of range/implausible} - This DTC describes an invalid value arriving at the motor controller.
Even though it specifies a clear symptom, it does not yet contain information about the origin of that faulty behavior.
In stark contrast to what some people may think, diagnostic trouble codes do not directly indicate the necessary actions for repair.
After analyzing the set of trouble codes, a technician might resort to differential diagnostics using off-board tools.
In case of the demostrated DTC, this might include validating the correct supply voltage to the sensor itself; 
validating the supply voltage to the motor control unit; validating the produced fuel pressure by utilizing an external pressure gauge; and validating the ohmic resistance of the signal line with a multi meter.
This differential diagnostic search tree is used and ordered to rule out one possible problem after another.
Depending on the read-out trouble codes, such a search tree might take up to hours or even days, resulting not only in a lenghty waiting process for the workshop's customer, but also hassle for the auto shop itself in terms of inner shop logistics.
In this very case, one single oscilloscope measurement measuring the signal on the line at the sensor connector and the motor controller connector could have been used to calculate the signal's transfer function $H(s)$ to rule out a whole plathora of electrical problems.
One of the reasons why this isn't done in auto shops lies in the mathematical complexity of this action itself as well as in non-existing data to compare the resulting transfer function to.
Using online services to store data to compare such values to and to calculate these functions out of time domain recordings could be an approach that takes the responsibility for the intellectual overhead.
By using a USB oscilloscope connected to a GUI program which presents the auto shop worker with clear instructions on how to attach the oscilloscope channels to specific signal lines in the car, automatically recording the data and sending it to a service is the key approach of this project. 
This proposed approach leads to multiple questions that need to be answered:
\begin{itemize}
  \item How can a GUI program support the technician workflow the best?
  \item How can a low-cost yet rugged oscilloscope be manufactured in order to be rolled-out to the majority of workshops without introducing financial or usability barriers?
  \item How can ever-improving analysis models be provided while simultaneously feeding back labled training data to algorithm developers?
  \item Which data privacy aspects need to be considered for providing training data with high significance without opening malicious attack vectors?
  \item How can training data be efficiently distributed to enable model developers to generate significant analysis models for the ever-growing variety of cars?
\end{itemize}

\subsection{Proof-of-Concepts}
In order to address these questions, the project \textbf{autowerkstatt4null} aims to run proof of concept experiments.
Goals of these experiments are answering to the general feasibility as well as to discover more detailed questions.
The proposed system can be compartmentalized into the following fields:
\begin{enumerate}
  \item Hardware
  \item GUI usability
  \item Data transport and service provisioning 
\end{enumerate}

\subsubsection{Data Transport and Provisioning}
In order for a produtive deployment of the proposed system, it must be figured out in what way acquired data from the workshop can be shared as a privacy preserving datum in a larger training data set, while also maintaining relevant meta information.
Not only does training data need to be shared but also models which were generated on this very training data set.
In succession of generating training data set and distilling analytical models, these models have to be distributed or provided as a service.
Analytical models can be conceptualized as functions $Analysis = Model(Measurement)$.
\begin{quote}
\textbf{Analytical model.}  
Let $x := (x_t)_{t=0}^{T}$ denote a $d$-dimensional time series with $x_t \in \mathbb{R}^d$ (the \emph{measurement}).  
An analytical model $f_\theta$ maps this measurement to an analysis object:
\[
f_\theta : (\mathbb{R}^d)^{T+1} \to \mathcal{A}, \qquad a = f_\theta(x).
\]
Here, $\mathcal{A}$ denotes the space of possible analyses (e.g., classes, scores, or structured diagnostic reports).  
Optionally, metadata $m$ may be included: $a = f_\theta(x, m)$.  
In probabilistic form one often writes $p_\theta(a \mid x)$ and derives, for example, the most likely analysis 
$a^\star = \arg\max_a p_\theta(a \mid x)$.
\end{quote}
This function needs to be evaluated on a machine.
Since it is expected that a model is not generated by a workshop, in a workshop, but rather by a signal conditioning expert that has no connection to the workshop, the measurement to be analyzed and the model to analyze with are stored on different machines.
This means that either the measurement data has to be sent to another machine on which the service function is available or the service function has to be transfered to the workshop machine prior to the actual diagnostic work. \hfill \\
In the demonstrator application different ideas to tackle this challenge shall be explored. 

\subsubsection{GUI Usability}
While developing a new diagnostic system, the ease of use can be a contributor to a project's success.
Auto-shop-technicians face an ever-increasing time- and efficiency-pressure, hence an efficient and effective user interface to all use-cases, need to pass practical application tests inside real auto shops.
Furthermore, it must be made sure that the provided user-software is designed to be extended and forked without the necessity of refactoring.
Therefore, the \textbf{autowerkstatt4null} project aims at an open-source-development for the demonstrator software.

\subsubsection{Hardware}
Automotive-oscilloscopes have been used for diagnostics since the 1970s already.
The market is saturated, meaning that there are multiple options for each use-case available to the general auto shop.
Due to this, an evaluation of existing automotive-oscilloscopes shall be performed in order to find out whether or not available solutions meet the requirements of sturdiness, cost-benefit-ratio, and possibilities of intergration. \\

By proving a concept of intergration of these three layers, the project will outline a possible foundation for a next-generation off-board-diagnostics solution intergrating state-of-the-art electrical signal analysis and machine learning/artificial-intelligence capabilities.

\subsection{Demonstrator}
\begin{figure}[ht]
  \centering
  \includegraphics[width=0.9\linewidth]{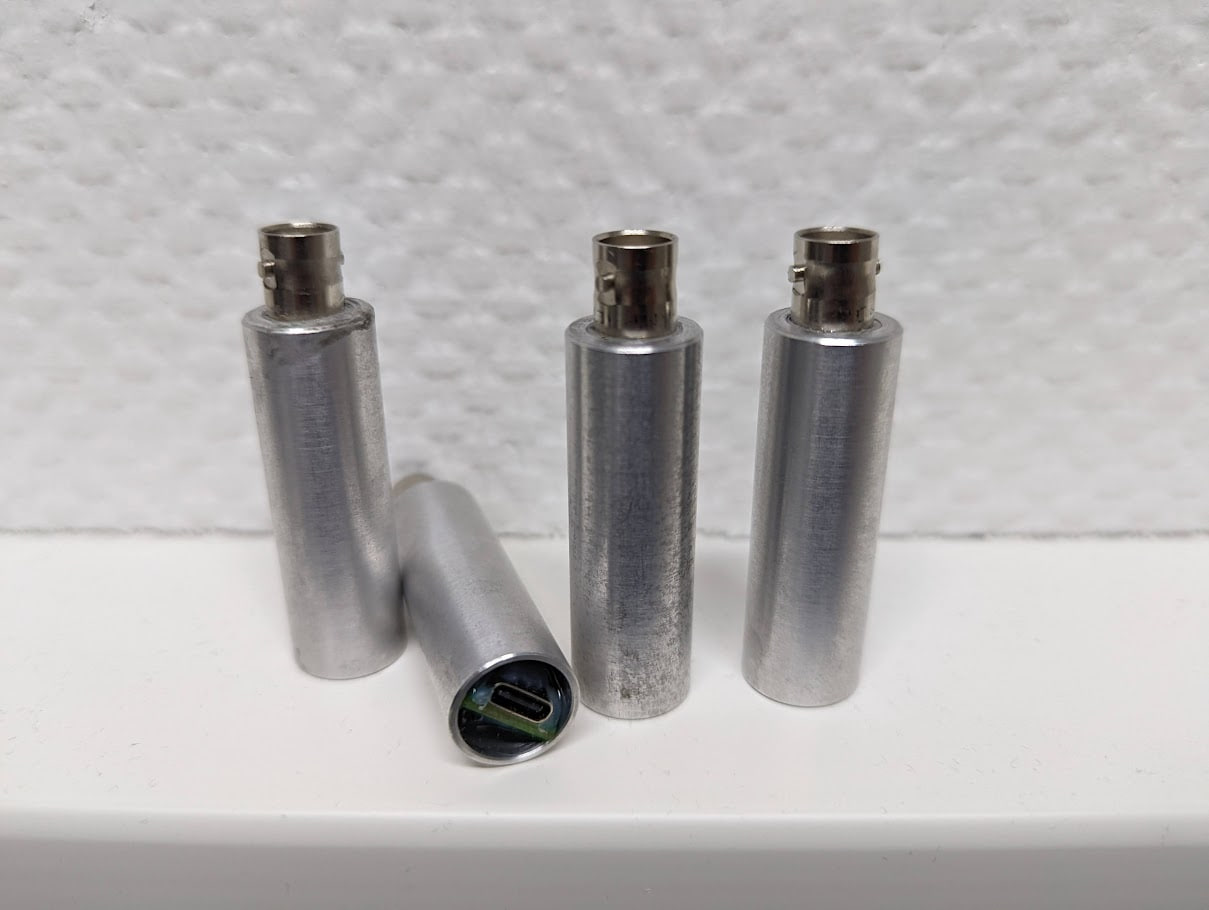}
  \caption{Early-Scope measurement prototype}
  \label{fig:early-scope}
\end{figure}

\begin{figure}[ht]
  \centering
  \includegraphics[width=0.9\linewidth]{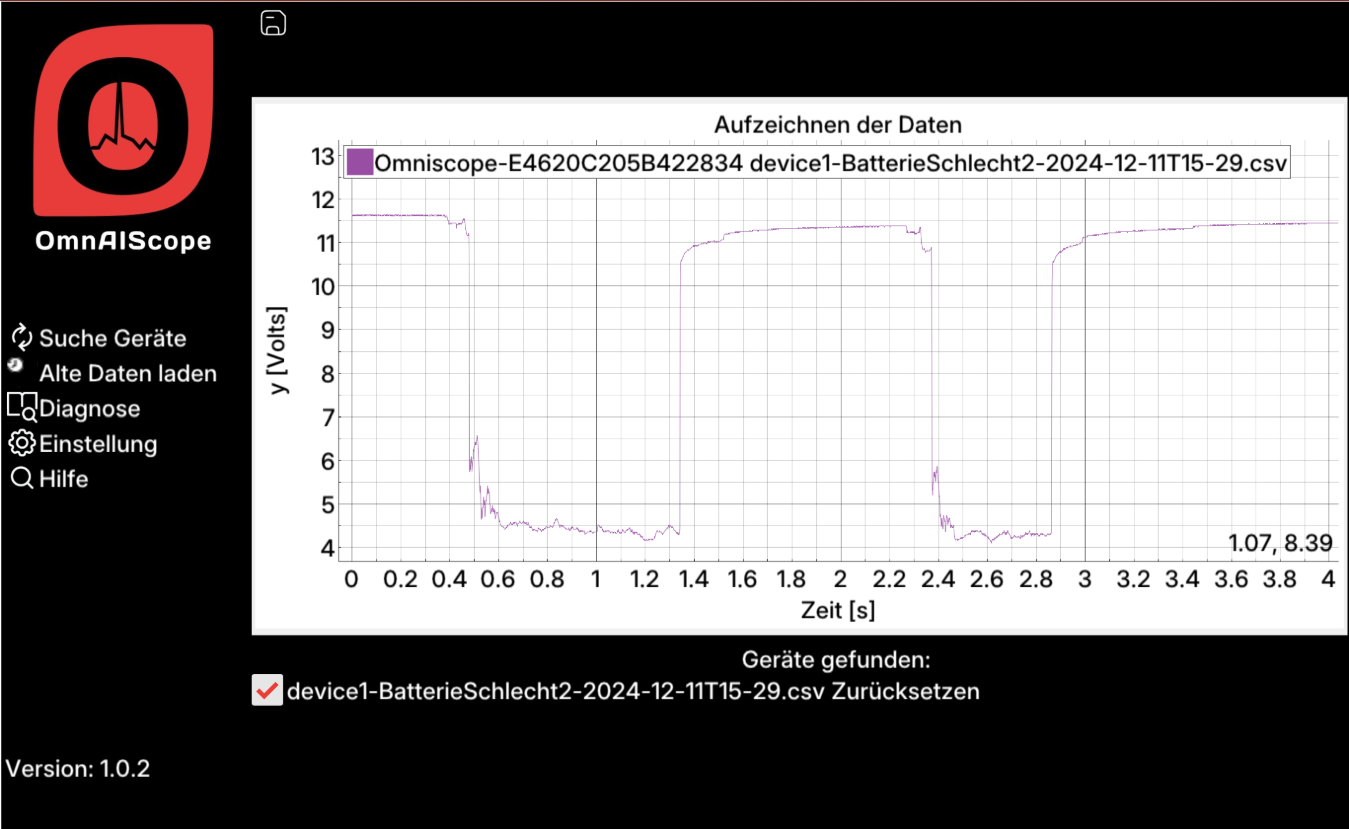}
  \caption{Prototype GUI: OmnAIView}
  \label{fig:gui}
\end{figure}

The current demonstrator consists of six primary components:
\begin{itemize}
  \item A Latop
  \item Installed OmnAIView software
  \item Installed VCDS Software
  \item A VCDS diagnostic tool
  \item Four OmnAIScopes
  \item Lab-Connectors (e.g., Banana-to-BNC adapters)
  \item A formal contract
  \item Instructional materials
\end{itemize}
In accordance with the proposed workflow for car diagnostics, the process begins with connecting the vehicle to a VCDS device, which performs a standard OBD scan. 
As described in the Introduction, this step typically yields vague and sometimes insufficient DTCs, which are then displayed and recorded on the local laptop.

To improve diagnostic specificity, USB-based oscilloscopes are connected to various electrical sensors in the vehicle using Banana-to-BNC adapters. 
The measurement prototype setup is shown in Figure~\ref{fig:early-scope}.
Each scope records time-varying voltages from a specific sensor. 

Since the microcontrollers themselves have no inherent notion of absolute time, synchronization relies on the USB protocol clock, which defines when one bit ends and the next begins. 
Therefore multiple USB-scopes have to be connected to the same root USB hub to be syncronized by the USB start-of-frame signals, as different hubs each maintain their own clock domain.
The measurement computer runs a dedicated data receiver service that collects the signals, aligns them based on the USB clock, and forwards the aggregated multichannel data stream to the OmnAIView GUI for oscillographic visualization as displayed in Figure~\ref{fig:gui}.

Collected measurement data is transferred via HTTP from the local hub to a remote gateway-server. 
This forwards the data to the central training-data storage-server, for future analysis. 

The long-term vision involves utilizing this stored signal data to train machine learning models capable of classifying faults based on signal patterns. 
However, due to the current lack of sufficient training data in the THGA repository, real-time diagnostic feedback based on these models is not yet operational. 
This aligns directly with the discussion in Machine Learning for Signal Clustering-highlighting the need for a significant and diverse training dataset before meaningful fault classification can be achieved.

\section{Results}
\subsection{Technical Readiness}
The current demonstrator successfully proves the technical feasibility of a physical data driven diagnostic workflow for automotive workshops. 
During initial testing, the system exhibited data transmission from oscilloscopes to a locally running data server.
Data was visualized via the local interface, providing technicians with immediate feedback from sensor measurements.
Additionally, data could be labeled in the OmnAIView GUI and VCDS scans could be attached before being stored on the training-data server.

From a throughput perspective, signal samples from multiple oscilloscopes were consistently streamed and rendered without major delay, even in parallel operation of multiple scope channels. 

Through hands-on work with the demonstrator, several key insights have emerged. 
Most notably, the clearest value for technicians lies in the immediate reception and visualization of sensor data-bridging the current diagnostic gap left by standard OBD scans. 
While conventional tools like VCDS provide a static view of ECU status and DTCs, the live waveform monitoring provides an accessible dimension to vehicle diagnostics.

However, the current system still consists of individually developed components that require manual coordination. 
Moving forward, a critical step will be to unify these elements into a cohesive software architecture that abstracts complexity for the end user. 
Additionally, the project now has a solid foundation to expand into data analysis, including both classical signal processing and machine learning models. 
These models could detect fault patterns, classify sensor behavior, and support predictive diagnostics; ultimately enabling technicians to make faster and more accurate repair decisions.

\subsection{Usage and Feedback}
The demonstrator has been deployed in 51 workshop environments over a six-month field test period, 
all of which were personally deliverd and users were on-boarded in a day-long workshop.
While a hand full of workshops produced data on a daily basis, other shops were only using the system rather occasionally. 
The feedack regarding the scope hardware was rated good in terms of usability and durability, even though some shops mentioned the physical dimensions of the device being a little too compact for some employees.
The user-software itself got less satisfying feedback. 
Especially the GUI which was implemented using dear-imgui, struck the users as \textit{old-fashioned} and not intuitive in its usage.
Moreover, key analysis features such as configurable triggers and advanced signal analysis tools, were reported as missing or insufficiently developed.
Another challenge which was encountered during the testing periode was the situation, that some of the workshops didn't have stable internet connections on their shopfloor.
This lead to interrupted data-transmission to the servers, while -due to a bug in the implementation- displaying a \textit{successfully sent}-message to the user.
All-in-all the feedback from the workshops was satisfying to the project team.
However, several challenges remain:
\begin{itemize}
  \item The various tools (VCDS, OmnAIView, hub interface, ML services) have not yet been unified into a single software platform, which limits usability.
  \item Training data volume remains insufficient, restricting the machine learning module’s ability to generate actionable feedback.
\end{itemize}

\subsection{Gathered Data}
The gathered data is currently stored in a cephfs storage network located at the THGA in Bochum.
Right now there is no publicly available dataset, that was extracted.
It is aimed to publicize statistics of fault-codes, usage and access to training-data within a step. 

\section{Summary}
The \textbf{autowerkstatt4null} project has made substantial progress in developing an innovative off-board diagnostics ecosystem for independent automotive workshops. 
The proposed architecture integrates USB-based OmnAIScopes, a central data hub, and REST/WebSocket communication protocols, 
enabling secure and efficient data flow from vehicle sensors to cloud-based analysis servers. 
This framework facilitates live diagnostics through real-time waveform visualization via the OmnAIView software, 
offering improved fault detection specificity compared to traditional OBD systems. 
Field tests in 51 workshops over a six-month period validated the system's technical feasibility, 
demonstrating data transmission and low-latency signal rendering, which provides technicians with immediate diagnostic insights. 
While challenges such as software unification and the need for larger training datasets for AI-driven diagnostics remain, 
these accomplishments provide a strong foundation for advancing next-generation automotive diagnostics.

\section{Outlook}
Based on the \textbf{autowerkstatt4null} project result, researchers, engineers, and application partners are convinced that the proposed, 
implemented, and tested system shows potential to advance workshop-diagnostics in terms of specificity and ease of use.
Given these first experimental results, two approaches will be taken to progress with further developments. 

First: a clear road map of future monetization of the project and its results has to be laid out in order to provide an environment for sustainable product development. 
Even though public and private research funding is, or might be available the Auto-Intern GmbH is determined to release an early derivative of the project's hardware to the general public market. 
This will lead to further improvements on the actual scope hardware and firmware. 
Anyhow, due to the nature of this endeavour, and the substantial investment, thus a necessary amortization, hard- and firmware development by the Auto-Intern GmbH will be progressing in closed source. 

Besides of that, three branches of development shall be advanced with as much public and open-source contribution as possible. 
Use-cases; application-software; as well as storage- and analysis- infrastructure will be organized in independent subgroups, 
the progress of which can be followed-up on under \url{https://github.com/omnai-project}.

\section{Ackowledgement}
This project was made possible with financial support of Auto-Intern GmbH, LMIS AG, Dekra, Verglöst, nabla-B UG, DMT-LB and its subsidiary THGA.
This work has been supported by the Bundesministerium für Wirtschaft und Klimaschutz (BMWK) project number 68GX21005E and the \href{https://gaia-x.eu/}{GAIA-X Project}.
A detailed list of contributions can be obtained from Lukas Jakubczyk at the University for Applied Sciences THGA in Bochum; \href{mailto:lukas.jakubczyk@thga.de}{lukas.jakubczyk@thga.de}, \href{https://www.thga.de/forschung/maschinenbau-und-materialwissenschaften/labore/prolab-produkt-produktion}{PROLAB für Produkt + Produktion}, \href{https://maps.app.goo.gl/tQA3Qfn5f6gGqMHg8}{Technische Hochschule Georg Agricola, Herner Str. 45, DE-44787 Bochum}.
We appreciate the support of all connected Car-Workshops and associated partners. 
This paper was edited and is maintened by Meihui Huang (\zh{黄美慧}). 
To leave comments or get a connection to the project, please use the \href{https://github.com/nabla-B/paper_aw4null-overview/issues}{github-issue tracker} or send a mail to \href{mailto:meihui@nabla-b.engineering}{meihui@nabla-b.engineering}

\bibliographystyle{IEEEtran}
\bibliography{src/references}
\end{document}